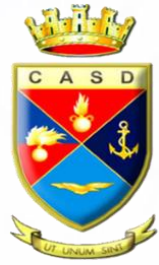

# DOCUMENTO FINALE DI SINTESI

Dibattito sulla Difesa e Sicurezza Sistemica

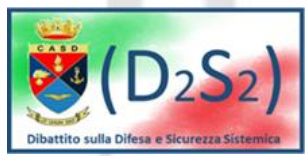

# PREFAZIONE

Nel riconfigurarsi quale Scuola Superiore a Ordinamento Speciale della Difesa, il Centro Alti Studi per la Difesa si è voluto porre quale interlocutore comprimario, affiancato dal Ministero degli Affari Esteri e della Cooperazione Internazionale e dal Dipartimento delle Informazioni per la Sicurezza, di un dialogo che fosse interforze ed inter-istituzionale, coinvolgendo anche il mondo dell'università e della ricerca. Da tale idea è nato il Dibattito sulla Difesa e Sicurezza Sistemica, che ha visto impegnati i co-proponenti dal luglio 2020 al dicembre 2021, giungendo infine, nel gennaio 2022, a conclusione.

L'iniziativa ha permesso una raccolta di contributi di pensiero di qualità, che si sono focalizzati sulle varie accezioni di cosa significhino termini quali *difesa* e *sicurezza* in un mondo che sta venendo ridefinito dal protrarsi della situazione pandemica. Sempre più evidentemente, le Forze Armate e il comparto dell'Intelligence si trovano ad operare in uno scenario di sfide mutate e mutevoli, ove solo una risposta di sistema intrinsecamente affiatata e decisa può garantire i risultati eccellenti pretesi. È dunque innegabile che il contributo di tutti i partecipanti, siano essi autori singoli o gruppi, appartenenti alle istituzioni, al mondo dell'università e della ricerca, a quello imprenditoriale, o in veste di privati cittadini, abbia permesso di elaborare un volume finale in cui le prospettive di opportunità e rilancio si sposano con un'analisi attenta e puntuale delle sfide presenti e potenziali.

È questo, a mio parere, un lavoro che esprime le intrinseche potenzialità di un Sistema Paese in grado di affrontare scenari e difficoltà provenienti dallo scacchiere internazionale con la sicurezza di una formazione eccellente ed una capacità di risposta sempre pronta a cogliere, nelle diversità dei comparti coinvolti, la propria forza.

*Ammiraglio di Divisione Giacinto Ottaviani*
*Presidente del Centro Alti Studi per la Difesa*

INFODEMIA E PANDEMIA: LA *COGNITIVE WARFARE* AI TEMPI DEL SARS-COV-2 —
Francesco Saverio Bucci[*] & Matteo Cristofaro[**] & Pier Luigi Giardino[***]

*Con la comparsa del SARS-CoV-2 si sono diffuse due epidemie: una sanitaria ed una informativa. Il virus ha generato un'infodemia senza precedenti, contribuendo ad instaurare un clima di forte incertezza. L'informazione, massiva e ridondante, è stata efficace vettrice anche di propaganda, su scala globale, ad opera di attori statali e non statali. Un vero e proprio atto ostile in cui non è prevista violenza fisica, ma gestione sistematica dell'informazione mediante la manipolazione della sfera cognitiva. Per questo motivo, l'adozione di un atteggiamento di analisi critica attiva da parte di cittadini ed istituzioni e altresì l'implementazione di politiche comuni a livello internazionale potrebbero, indubbiamente, facilitare il contrasto alla guerra cognitiva (c.d.* cognitive warfare*) e dunque limitarne gli effetti disastrosi.*

La diffusione globale del SARS-CoV-2 ha originato un'infodemia[960,961], una circolazione indiscriminata di notizie affliggendo anche la comunicazione istituzionale che, pur ispirandosi al principio di apportare ordine in un sistema informativo emergenziale in totale disordine, è stata travolta dagli eventi, registrando indugi, distorsioni e fughe di notizie[962,963].

---

[*] Francesco Saverio Bucci – Su designazione del Capo di Stato Maggiore della Difesa, completando la sua presenza formativa e di tutoraggio presso il CASD, ha frequentato la 65ª Sessione dell'IASD. Ha partecipato a working group e harmonization conference promosse dal NATO SHAPE. È autore di lavori inerenti al ruolo dell'*intelligence* a sostegno della lotta al terrorismo internazionale, a supporto della difesa da armamenti non convenzionali/di distruzione di massa e nell'evoluzione nella gestione della conflittualità asimmetrica e dei fattori cognitivi nei processi di analisi di *intelligence*

[**] Matteo Cristofaro – PhD, Assegnista di ricerca in Management e docente presso l'Università degli Studi di Roma "Tor Vergata", Dipartimento di Management e Diritto.

[***] Pier Luigi Giardino – Laureando magistrale in Business Administration presso l'Università degli Studi di Roma "Tor Vergata".

[960] Atefeh Vaezi e Shaghayegh Haghjooy Javanmard. "Infodemic and risk communication in the era of CoV-19," *Advanced Biomedical Research* 9 (2020): 10.

[961] "Coronavirus disease (COVID-19) Situation Report - 163", World Health Organization, ultimo accesso 28 maggio 2021, https://apps.who.int/iris/bitstream/handle/10665/332971/nCoVsitrep01Jul2020-eng.pdf.

[962] Marco Centorrino, "Infodemia e comdemia: la comunicazione istituzionale e la sfida del COVID-19". Humanities 9, no. 2 (2020): 1-18.

[963] Lelio Alfonso e Gianluca Comin. *#ZONAROSSA. Il Covid-19 tra infodemia e comunicazione*. Guerini e Associati, 2020.

È ormai acclarato che la diffusione di attacchi informativi esploda contemporaneamente a situazioni di crisi[964], con il potenziale di accrescere le paure e l'ansia dei cittadini e ostacolare un'efficace gestione da parte di governi e istituzioni. Se negli anni Novanta Mattelart scriveva "l'informazione è qualcosa che serve a muovere guerra"[965], oggi è sempre più chiaro che l'informazione costituisce un'arma a tutti gli effetti e come tale vada integrata nella strategia di Difesa nazionale.

La Relazione sulla Politica dell'Informazione per la Sicurezza del 2020[966] ha evidenziato che i rapporti economici e politici sono stati drammaticamente sconvolti dalla diffusione del SARS-CoV-2. Oltre agli effetti sanitari direttamente collegati alla pandemia, è stata registrata una elevatissima produzione di notizie false e narrazioni allarmistiche, sfociate in un'infodemia di difficile discernimento per la collettività. A livello sociale, le manipolazioni informative hanno diffuso informazioni distorte e fabbricate *ad hoc*, sfruttando le vulnerabilità cognitive dell'essere umano, predisponendolo ad accettare informazioni non validate[967].

Per comprendere il fenomeno di un'informazione viziata è necessario delineare il confine tra le tipologie di manipolazione informativa: *i) misinformazione*, cioè la diffusione falsa informazione senza un intento preciso di causare danni a qualcosa o qualcuno, *ii) disinformazione*, intesa come la diffusione di falsa informazione con l'intento di causare danni a qualcosa o qualcuno, e *iii) malinformazione*, identificata come la diffusione di informazione 'vera' con l'intento di causare danni a qualcosa o qualcuno[968,969].

---

[964] Manuel Castells, *The information age*. Blackwell Publishers, 1996.

[965] Armand Mattelart, *Mapping world communication: War, progress, culture*. Univesity of Minnesota Press, 1994.

[966] "Relazione sulla politica dell'informazione per la sicurezza 2020", Sistema di Informazione per la Sicurezza della Repubblica, ultimo accesso 28 maggio 2021, https://sicurezzanazionale.gov.it/sisr.nsf/wp-content/uploads/2021/02/RELAZIONE-ANNUALE-2020.pdf.

[967] Emilio Ferrara, Stefano Cresci e Luca Luceri. "Misinformation, manipulation, and abuse on social media in the era of COVID-19," *Journal of Computational Social Science* 3, no. 2 (2020): 271-277.

[968] Karen Santos-D'amorim e Májory K. Fernandes de Oliveira Miranda. "Misinformation, Disinformation, and Malinformation: Clarifying the definitions and examples in disinfodemic times," *Revista Eletrônica de Biblioteconomia e Ciência da Informação* 26 (2021): 1-23.

[969] Darrin Baines e R. J. Elliott. "Defining misinformation, disinformation and malinformation: An urgent need for clarity during the COVID-19 infodemic," *Discussion Papers* 20 (2020).

L'attuazione di campagne di disinformazione e malinformazione richiede una profonda comprensione delle dinamiche sociali e cognitive ed una consapevolezza temporale e metodologica al fine di poter meglio sfruttare le vulnerabilità offerte dalle diverse circostanze. Le operazioni di influenza ostile e sfruttamento della cognizione umana sono avvantaggiate dall'introduzione di nuovi strumenti tecnologici, come l'intelligenza artificiale, che forniscono ai propagandisti metodologie più precise, complesse e imprevedibili. In particolare, il sempre più diffuso uso di *social network* e piattaforme di messaggistica istantanea ha contribuito alla creazione di terreno fertile per innescare, e/o rafforzare, azioni geopolitiche attraverso una massiccia attività di influenza cognitiva[970,971]. Nella fattispecie, durante la pandemia si sono registrate attività manipolative da parte di taluni Paesi con l'intento di creare divisioni tra le nazioni occidentali e minare la fiducia nei relativi governi[972,973]; una vera e propria proliferazione di atti ostili di *cognitive warfare*.

La *cognitive warfare* è da considerare come un conflitto ibrido organizzato, in cui soggetti si confrontano sulla capacità di produrre, mettere in relazione ed eludere elementi di conoscenza, al fine di costruire rappresentazioni mentali generalizzate nell'opinione pubblica con la finalità di orientare emozioni, ragionamenti e comportamenti dei soggetti[974]. Appare chiara, da tale definizione, la cospicuità della *cognitive warfare* nella quale tutti i soggetti partecipano, per lo più inavvertitamente, tramite l'elaborazione, lo sviluppo e la diffusione delle informazioni[975], incrementando la magnitudo degli effetti. Nello specifico, la *cognitive warfare* si serve di un approccio multidisciplinare che indaga i processi conoscitivi e comunicativi, servendosi di numerose metodologie e strategie, spesso utilizzate in modo coordinato.

---

[970] Mikhail D. Suslov, ""Crimea is Ours!" Russian popular geopolitics in the new media age," *Eurasian Geography andEeconomics* 55, no. 6 (2014): 588-609.

[971] Konstantin Platonov e Kirill Svetlov. "Conspiracy Theories Dissemination on SNS Vkontakte: COVID-19 Case," In *International Conference on Electronic Governance and Open Society: Challenges in Eurasia*, pp. 322-335. Springer, Cham, 2020.

[972] Costanza Sciubba Caniglia, "Signs of a new world order: Italy as the COVID-19 disinformation battlefield," *Harvard Kennedy School Misinformation Review* 1, no. 3 (2020).

[973] Robin Emmott, "Russia, China sow disinformation to undermine trust in Western vaccines: EU", *Reuters*, 28 aprile 2021, https://www.reuters.com/world/china/russia-china-sow-disinformation-undermine-trust-western-vaccines-eu-report-says-2021-04-28/.

[974] Stuart A. Green, "*Cognitive Warfare*" (MA diss., *United States Joint Military Intelligence College*, 2008).

[975] François du Cluzel, *Cognitive Warfare.* NATO Innovation Hub, 2020.

Al fine di cristallizzare le dinamiche della *cognitive warfare*, è necessario approfondire alcuni aspetti partendo dal concetto di razionalità limitata. Nel 1947, Herbert Simon formulò l'ipotesi che l'essere umano è limitato nel tentativo di agire razionalmente a causa dalla sua: *i)* ristretta capacità di calcolo, *ii)* impossibilità di accesso a tutte le informazioni, e *iii)* intrinseca biologia[976]. Secondo Simon, l'essere umano possiede una base cognitiva che non consente di prendere decisioni libere da errori cognitivi e deviazioni irrazionali; in particolare, la cognizione dell'essere umano è soggetta a due sub-categorie di errori cognitivi: le euristiche – scorciatoie mentali utilizzate in situazioni ad alta complessità per produrre risposte in tempi brevi – e le trappole decisionali – deviazioni della razionalità che influenzano le decisioni dell'essere umano. In questo scenario, due delle maggiori variabili sfruttate per mettere in atto o contrastare campagne di *cognitive warfare* sono gli errori cognitivi dell'*information overload* e l'euristica affettiva[977].

L'*information overload* è l'errore cognitivo per il quale gli individui che ricevono un volume di dati perdono l'abilità di filtrare e classificare gli input ricevuti in base alla loro rilevanza, veridicità ed utilità. La conseguenza è produzione di *output* informativi distorti e la condivisione di notizie false, decontestualizzate o parziali[978].

L'euristica affettiva[979] fornisce una precisa descrizione di quelli che sono gli antecedenti emotivi che influenzano un individuo nell'adottare decisioni. In sintesi, sia lo stato emotivo iniziale dell'individuo che la carica emozionale delle informazioni ricevute contribuiscono a una valutazione dell'ambiente decisionale nel quale il soggetto è immerso[980]. È utile rilevare come queste euristiche comportino importanti limiti e come i propagandisti possano impiegarle a loro vantaggio.

---

[976] Herbert A. Simon, *Administrative behavior*. The Free Press, 1947.

[977] Daniel Kahneman, "Maps of bounded rationality: Psychology for behavioral economics," *American Economic Review* 93, no. 5 (2003): 1449-1475.

[978] Angela Edmunds e Anne Morris. "The problem of information overload in business organisations: a review of the literature," *International Journal of Information Management* 20, no. 1 (2000): 17-28.

[979] Matteo Cristofaro, ""I feel and think, therefore I am": An Affect-Cognitive Theory of management decisions," *European Management Journal* 38, no. 2 (2020): 344-355.

A contrasto di azioni manipolative e di propaganda, risulta efficace l'adozione di un atteggiamento di analisi critica attiva che richiede l'adozione di un approccio di controllo della credibilità maggiormente sistematico e consapevole dei meccanismi di manipolazione. I criteri da considerare quando si fa riferimento ai messaggi scambiati mediante qualunque canale informativo, verticale o orizzontale, sono l'accuratezza, la coerenza, la completezza, la controllabilità e l'attendibilità delle fonti[981,982]. L'analisi critica attiva significa quindi consapevolezza delle tecniche persuasive utilizzate nella diffusione dell'informazione, dei rischi derivanti dalle decisioni influenzate dalle manipolazioni, del ruolo delle emozioni nel mediare il pensiero analitico nelle decisioni e dell'esigenza di diffondere competenze legate alla ricerca attiva di fonti informative alternative e alla loro valutazione[983,984,985].

Se i virus informativi, similmente ai biologici, possono generare severi livelli di destabilizzazione è necessario contrastare le infodemie proprio come si reagisce alle malattie, comprendendo sviluppi, diffusioni, propagazione e *cluster* che potrebbero essere targettizzati. Indispensabile rafforzare un sistema in termini di *media* e *digital awarness* per garantire un'opera critica dei fenomeni e una tutela dai virus informativi.

---

[980] Matteo Cristofaro, "The role of affect in management decisions: A systematic review," *European Management Journal* 37, no. 1 (2019): 6-17.

[981] Timothy P. McGeehan, "Countering Russian Disinformation," *Parameters* 48, no. 1 (2018): 49-57.

[982] Stephan Lewandowsky e Sander Van Der Linden. "Countering misinformation and fake news through inoculation and prebunking," *European Review of Social Psychology* (2021): 1-38.

[983] John J. Doherty, Mary Anne Hansen e Kathryn K. Kaya. "Teaching information skills in the information age: the need for critical thinking," *Library Philosophy and Practice* 1, no. 2 (1999): 1-9.

[984] Robert Duron, Barbara Limbach e Wendy Waugh. "Critical thinking framework for any discipline," *International Journal of Teaching and Learning in Higher Education* 17, no. 2 (2006): 160-166.

[985] Maryellen Allen, "Promoting critical thinking skills in online information literacy instruction using a constructivist approach," *College & Undergraduate Libraries* 15, no. 1-2 (2008): 21-38.

Come emerso dalla relazione al Parlamento, avvenuta nel maggio del 2020[986], dell'On. Volpi, al tempo Presidente del CoPaSIR, l'Italia è stata obiettivo di intense attività di *cognitive warfare* mirate alla diffusione, in maniera coordinata e continuativa, di notizie fuorvianti riguardanti vaccini, rimedi terapeutici e strumenti diagnostici. È evidente che l'attività infodemica si è collocata in un contesto geopolitico globale nel quale il SARS-CoV-2 ha rappresentato lo scenario migliore per permettere ad alcuni Paesi – in particolare Russia e Cina – di mostrare un'ipotetica maggiore efficienza rispetto alle controparti occidentali, nel contrasto alla propagazione del virus, collocando l'Italia come un obiettivo di primario interesse per la diffusione di notizie allo scopo di alimentare l'incertezza, screditare la reputazione statuale e creare un clima di sfiducia[987,988,989,990].

In particolare, tali Paesi si sono adattati alle nuove sfide nell'infosfera[991], manifestando velocità, aggressività e determinazione nel raggiungimento dei loro obiettivi. La ragione di questa *escalation* è la debole reazione alla nuova tipologia di minaccia e la mancanza di una ferma e sistematica controffensiva. Il Governo ha il compito di affrontare la *cognitive warfare* come minaccia in espansione, adeguando una reazione proporzionale, aumentando la consapevolezza informativa, integrando le capacità operative e finalità strategiche da essa scaturite nelle agende politiche e predisponendo quindi strumenti necessari al suo contrasto.

---

[986] "L'allarme del Copasir: "Contro il nostro Paese fake news virali sull'epidemia di coronavirus"", *RaiNews*, 26 maggio 2020, http://www.rainews.it/dl/rainews/articoli/allarme-del-Copasir-contro-il-nostro-Paese-fake-news-virali-su-epidemia-di-coronavirus-6fe475b0-e465-4507-9917-6da21785e9ec.html.

[987] Rose Bernard, Gemma Bowsher, Richard Sullivan e Fawzia Gibson-Fall, "Disinformation and epidemics: Anticipating the next phase of biowarfare," *Health Security* 19, no. 1 (2021): 3-12.

[988] Radu Magdin, "Disinformation campaigns in the European Union: Lessons learned from the 2019 European Elections and 2020 Covid-19 infodemic in Romania," *Romanian Journal of European Affairs*, 20(2), (2020): 49-61.

[989] Alec Luhn, "Ex-Soviet countries on front line of Russia's media war with the west", The Guardian, ultimo accesso 15 aprile 2021, https://www.theguardian.com/world/2015/jan/06/-sp-ex-soviet-countries-front-line-russia-media-propaganda-war-west.

[990] Benjamin Strick, Olga Robinson e Shayan Sardarizadeh, "Coronavirus: Inside the pro-China network targeting the US, Hong Kong and an exiled tycoon", BBC News, ultimo accesso 15 aprile 2021, https://www.bbc.com/news/blogs-trending-52657434.

[991] Jakub Kalenský, "Infosphere: the need to reverse a losing trajectory", NATO Defense College Foundation, ultimo accesso 15 aprile 2021, https://www.natofoundation.org/game-changers-2020-dossier-responses-to-disinformation/.

La *cognitive warfare* rappresenta, quindi, una minaccia effettiva e reale alla stabilità politica, sociale ed economica nazionale, manifestandosi come un'arma efficace a tutti gli effetti e come tale da integrare nelle strategie di Difesa nazionale. Inoltre, essendo le campagne di *cognitive warfare* spesso dirette contro interi sistemi politici ed economici come Unione Europea e NATO, esse andrebbero combattute mettendo in atto politiche comuni di prevenzione e contrasto facendo sì che agendo in maniera sinergica si possano limitare i danni causati al mondo democratico.

Pertanto, si ritiene fondamentale la creazione e/o rafforzamento di strutture permanenti nazionali ed internazionali – su modello NATO Centres of Excellence (COEs) – che favoriscano collaborazione e prontezza nell'identificare e quantificare i rischi connessi alla *cognitive warfare* nel breve e nel medio-lungo termine, grazie anche alla stretta cooperazione con i vari Sistemi di informazione per la sicurezza, fornendo così supporto agli organi politici sia in materia decisionale che legislativa.

# NOTA DI REDAZIONE

Il Documento Finale di Sintesi è stato elaborato dalla dott.ssa Ginevra Fontana all'interno del progetto "Dibattito sulla Difesa e Sicurezza Sistemica", svoltosi da luglio 2020 a dicembre 2021. Le opinioni espresse nell'elaborato sono da imputarsi esclusivamente all'autrice e non rispecchiano l'opinione né la posizione del Ministero della Difesa, del Centro Alti Studi per la Difesa, né di alcuna loro componente. Tutte le risorse online alle quali si fa riferimento sono state consultate in ultima istanza in data 25 ottobre 2021.

Il 1° Quaderno Speciale è stato elaborato dalla dott.ssa Ginevra Fontana all'interno del progetto "Dibattito sulla Difesa e Sicurezza Sistemica", svoltosi da luglio 2020 a dicembre 2021. Le opinioni espresse nell'elaborato sono da imputarsi esclusivamente all'autrice e non rispecchiano l'opinione né la posizione del Ministero della Difesa, del Centro Alti Studi per la Difesa, né di alcuna loro componente. Tutte le risorse online alle quali si fa riferimento sono state consultate in ultima istanza nel dicembre 2020.

Le opinioni esposte nei contributi di pensiero, ricevuti e resi disponibili nell'ambito delle iniziative *Call for Papers #CASD2020* e *Call for Papers #CASD2021*, sono da imputarsi esclusivamente ai rispettivi autori e non rispecchiano l'opinione né la posizione del Ministero della Difesa, del Centro Alti Studi per la Difesa, né di alcuna loro componente.

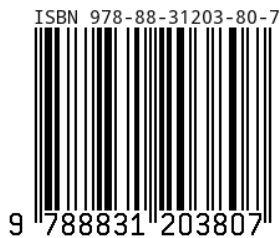